\documentclass[12pt]{article}
\usepackage{amsmath,amssymb}
\usepackage{mathtools}
\usepackage{type1cm}
\usepackage{graphicx}

\usepackage{color}
\definecolor{darkblue}{rgb}{0.15,0.35,0.55}
\definecolor{reddish}{rgb}{0.65, 0.2, 0.2}
\definecolor{phthaloblue}{rgb}{0.0, 0.06, 0.54}
\definecolor{bluscuro}{rgb}{0.15, 0.2, .85}
\definecolor{rossos}{cmyk}{0,1,1,0.55}
\definecolor{bluchiaro}{cmyk}{1,.3,0.,0.1}
\usepackage[bookmarks,linktocpage, colorlinks=true, plainpages = false, citecolor = darkblue,  linkcolor=darkblue, urlcolor = darkblue, filecolor =darkblue]{hyperref} 
\usepackage{lipsum}%dummy text
\usepackage{braket}

\usepackage{acro}
\DeclareAcronym{gr}{
    short=GR ,
    long=general relativity
}
\DeclareAcronym{uv}{
    short=UV ,
    long=ultraviolet
}
\DeclareAcronym{ir}{
    short=IR ,
    long=infrared
}
\DeclareAcronym{wdw}{
    short=WDW ,
    long=Wheeler-DeWitt
}
\DeclareAcronym{hl}{
    short=HL ,
    long=Ho\v{r}ava-Lifshitz
}
\DeclareAcronym{adm}{
    short=ADM ,
    long={Arnowitt, Deser and Misner}
}

\begin{document}
\setcounter{footnote}{0}
\begin{titlepage}
\begin{center}
\hfill YITP-24-54\\
\vskip .75in

%{\LARGE \bf
%Quantum perturbations arising \\[0.7cm]
%from no spacetime}
%\vskip .5in

{\Large \bf
No smooth spacetime: Exploring primordial \\[0.7cm]
perturbations in Lorentzian quantum cosmology }
\vskip .7in

{\large
  Hiroki Matsui$^{\rm (a)}$,}
\vskip 0.25in

$^{\rm (a)}${\em 
Center for Gravitational Physics and Quantum Information, 
Yukawa Institute for Theoretical Physics,
Kyoto University, 606-8502, Kyoto, Japan}

\end{center}

\vskip .5in

\begin{abstract}
Recent analysis of quantum cosmology has focused on the Lorentzian path integral formulations of both the no-boundary and tunneling proposals. However, it has been criticized that the wave function for linearized perturbations around a homogeneous and isotropic background leads to an inverse Gaussian distribution. This results in divergent correlation functions and cosmological inconsistencies. In this paper we explore these perturbation problems in Lorentzian quantum cosmology, focusing in particular on the quantum creation of the closed, flat, and open universe from no spacetime and the beginning of primordial inflation. We show that most quantum cosmological scenarios have serious perturbation problems. 
We also study the effects of trans-Planckian physics on quantum cosmology, using the generalized Corley-Jacobson dispersion as a case study of modified dispersion relations. Our findings indicate that resolving perturbation problems in Lorentzian quantum cosmology with modified dispersion relations remains a challenge. However, the perturbations can be Gaussian in the quantum creation of the flat or open universe within the confines of the saddle-point approximation and the generalized Corley-Jacobson dispersion. 
%This suggests that trans-Planckian physics or quantum gravity could potentially stabilize perturbations in quantum cosmology.

\end{abstract}

\end{titlepage}
\allowdisplaybreaks[1]
\tableofcontents

%%%%%%%%%%%%%%%%%%%%%%%%%%%%%%%%%%%%%%%%%%%%%%%%%%%%%%%%%%%
%%%%%%%%%%%%%%%%%%%%%%%%%%%%%%%%%%%%%%%%%%%%%%%%%%%%%%%%%%%
\section{Introduction}
%%%%%%%%%%%%%%%%%%%%%%%%%%%%%%%%%%%%%%%%%%%%%%%%%%%%%%%%%%%
%%%%%%%%%%%%%%%%%%%%%%%%%%%%%%%%%%%%%%%%%%%%%%%%%%%%%%%%%%%

Classical \ac{gr} does not explain the origin and evolution of the Planck-sized primordial universe, where quantum gravity effects become significant and quantum theory of gravity is essential.
Quantum cosmology aims to reveal the cosmological genesis based on an approach based on the quantum gravity framework. It introduces the concept of the so-called wave function of the universe, which is the wave-functional $\Psi[g,\phi]$ is defined on the space of all 3-geometries $g$ and matter field configurations $\phi$, called superspace.

The most famous models for the wave function $\Psi[g,\phi]$ are the no-boundary proposal~\cite{Hartle:1983ai} and the tunneling proposal~\cite{Vilenkin:1984wp}. These proposals have been recently investigated in the Lorentzian path integral approach in quantum gravity~\cite{Feldbrugge:2017kzv,DiazDorronsoro:2017hti}. 
This Lorentzian method rooted in the \ac{adm} formalism~\cite{Arnowitt:1962hi}, is a consistent path integral formulation of quantum gravity providing detailed insights into the no-boundary and tunneling wave function beyond \ac{gr}~\cite{Fanaras:2021awm,Narain:2021bff,Narain:2022msz,Ailiga:2023wzl, Lehners:2023yrj,Matsui:2023hei}.
However, there are concerns about their validity when considering linear perturbations around a homogeneous and isotropic background~\cite{Feldbrugge:2017fcc,Feldbrugge:2017mbc}. The wave function for these linearized perturbations takes an inverse Gaussian form, leading to diverging perturbation correlation functions. This implies that the anisotropy and inhomogeneity of spacetime due to these perturbations can not be suppressed, questioning the consistency of both the no-boundary and tunneling proposals with cosmological observations of an extremely homogeneous and isotropic universe.
In quantum cosmology, revealing the behavior of primordial perturbations is therefore a critical open issue.
%%%%%%%%%%%%%%%%%%%%%%%%%%%%%%%%%%%%%%%%%%
\footnote{
DeWitt's proposal~\cite{DeWitt:1967yk}, which posits the vanishing universe wave function at the Big Bang singularity, also faces challenges with perturbations in \ac{gr} and requires new quantum gravity theory such as \ac{hl} gravity~\cite{Matsui:2021yte,Martens:2022dtd}.}
%%%%%%%%%%%%%%%%%%%%%%%%%%%%%%%%%%%%%%%%%%

Many attempts to resolve such quantum perturbation problems in 
no-boundary and tunneling proposals based on the Lorentzian framework have been studied~\cite{DiazDorronsoro:2018wro,Feldbrugge:2018gin,Vilenkin:2018dch,Vilenkin:2018oja,Wang:2019spw,Bojowald:2018gdt,DiTucci:2019dji,DiTucci:2019bui,Halliwell:2018ejl,Bojowald:2020kob,Lehners:2021jmv}. Among these proposals, the most promising approach seems to be the modification of boundary conditions on the background and perturbations. However, as of now, there is no fully satisfactory solution that resolves all the issues.
In our previous work~\cite{Matsui:2022lfj}, we revisited this issue, extending these perturbation problems beyond \ac{gr} to include trans-Planckian physics, which modifies the dispersion relations for perturbations at wavelengths smaller than the Planck scale~\cite{Martin:2000xs,Brandenberger:2000wr,Niemeyer:2000eh,Martin:2002kt,Ashoorioon:2004vm,Ashoorioon:2011eg}. We demonstrated that the inverse Gaussian problem of perturbations in no-boundary and tunneling proposals is hard to overcome with the trans-Planckian physics modifying
the dispersion relation such as the generalized Corley-Jacobson dispersion relation~\cite{Corley:1996ar,Corley:1997pr} and the Unruh dispersion relation~\cite{Unruh:1994je}.

In this paper, we explore this approach in the contexts of flat, closed, and open universes. While the closed universe case has been extensively studied in the literature, the flat and open universe cases have not received as much attention and it is possible that the flat and open universe with non-trivial topology could emerge from nothing~\cite{Coule:1999wg,Linde:2004nz}. 
Therefore, this paper focuses on investigating the perturbation problem in Lorentzian quantum cosmology for flat and open universes incorporating trans-Planckian effects.
We also comment on the perturbation problems in the primordial inflation based on the Lorentzian quantum cosmology. This case was studied in the early work~\cite{DiTucci:2019xcr} and suggested that primordial inflation has such a problem. In this paper, we show that the quantum creation of closed, flat, and open universes from nothing, and the beginning of primordial inflation have serious perturbation problems. However, we demonstrate that the perturbations can be Gaussian in the quantum creation of flat and open universes within the confines of the saddle-point approximation and the generalized Corley-Jacobson dispersion.
In this sense, trans-Planckian physics or quantum gravity could potentially stabilize perturbations in quantum cosmology.

This paper is organized as follows: Section~\ref{sec:Lorentzian-quantum-cosmology} provides a brief overview of Lorentzian quantum cosmology, and explains how the no-boundary and tunneling proposal is achieved in the Lorentzian path integral formulation. We employ the saddle-point approximation and the Picard-Lefschetz theory. We can see that the saddle points associated with the no-boundary and tunneling proposals are complex, whereas those related to the quantum creation of flat and open universes are real. In Section~\ref{sec:Perturbation-analysis}, we discuss the perturbation problems in Lorentzian quantum cosmology and show that most quantum cosmological scenarios have serious perturbation issue. 
In Section~\ref{sec:TPP-modified}, we explore the trans-Planckian physics effects using the generalized Corley-Jacobson dispersion as a case study for modified dispersion relations at short distances. We show that the perturbation problems in Lorentzian quantum cosmology are hard to overcome with the trans-Planckian physics effects. 
Nevertheless, we demonstrate that the perturbation stability can be ensured in the quantum creation of flat and open universes within the confines of the saddle-point approximation and the generalized Corley-Jacobson dispersion.
Finally, in Section~\ref{sec:conclusions}, we summarize our conclusions.

%%%%%%%%%%%%%%%%%%%%%%%%%%%%%%%%%%%%%%%%%%%%%%%%%%%%%%%%%%%
%%%%%%%%%%%%%%%%%%%%%%%%%%%%%%%%%%%%%%%%%%%%%%%%%%%%%%%%%%%
\section{Lorentzian quantum cosmology}
\label{sec:Lorentzian-quantum-cosmology}
%%%%%%%%%%%%%%%%%%%%%%%%%%%%%%%%%%%%%%%%%%%%%%%%%%%%%%%%%%%
%%%%%%%%%%%%%%%%%%%%%%%%%%%%%%%%%%%%%%%%%%%%%%%%%%%%%%%%%%%

In this section, we will review 
the no-boundary and tunneling proposals based on Lorentzian quantum cosmology. 
The gravitational transition amplitude from the initial state $g_{i}$ to the final state
$g_{f}$ can be expressed by the gravitational path integral, 
\begin{equation}
G[g_{f};g_{i}]= \int_{{\cal M}}\mathcal{D}g_{\mu\nu}
~ \exp \left(\frac{i}{\hbar}S[g_{\mu\nu}]\right) ~,
\end{equation}
where $S[g_{\mu\nu}]$ is the Einstein-Hilbert action in \ac{gr}.
In general, the function $G[g_{f};g_{i}]$ serves as a Green’s function for the Wheeler-DeWitt equation, as discussed in~\cite{Teitelboim:1981ua}. If the geometries have a single boundary of $g_{f}$, then $G[g_{f};g_{i}]$ acts as a solution to the Wheeler-DeWitt equation, and the gravitational path integral defines the wave function of the universe. The Einstein-Hilbert action 
with a positive cosmological constant $\Lambda$ and a boundary term 
is written as
\begin{equation}
S[g_{\mu\nu}] = \frac{1}{2}\int_{\cal M} \mathrm{d}^4x \sqrt{-g} \left( R - 2 \Lambda\right) + \int_{\cal \partial M} \mathrm{d}^3y \sqrt{g^{(3)}} \mathcal{K}\,,
\end{equation} 
where we take the Planck mass unit with $M_{\rm Pl}=1/\sqrt{8\pi G}=1$.
The second term is known as the Gibbons-Hawking-York boundary term, involving the 3-metric $g^{(3)}_{ij}$ and the trace of the boundary's extrinsic curvature $\mathcal{K}$ on $\partial {\cal M}$. In gravitational path integrals, a commonly used method is the functional approach based on the Euclidean metric $g^{\rm E}_{\mu\nu}$. The Hartle-Hawking no-boundary proposal is a famous application of this approach~\cite{Hartle:1983ai}. Initially, this proposal suggested that the wave function of the universe could be represented by a path integral over all compact Euclidean geometries, characterized by a 3-dimensional boundary-only geometry. This proposal elegantly explains the quantum birth of the universe but has been criticized so far for various technical reasons since the Euclidean formulation of gravity is considered to be problematic~\cite{Gibbons:1978ac}.

Beyond the Euclidean formulation, it was suggested to take the path integral along the steepest descent paths in complex metrics. In this approach, it is not necessary to start with the Euclidean or Lorentzian metrics. Instead, considering $g_{\mu\nu}$ as complex, the integral is carried out along contours where the real part of the action increases most rapidly. However, in reality, such contours are not unique, bringing ambiguity between the Hartle-Hawking no-boundary proposal and Vilenkin's tunneling proposal~\cite{Halliwell:1988ik}. More recently, the no-boundary and tunneling proposals in minisuperspace quantum cosmology have been investigated by the Lorentzian path integral formulation~\cite{Feldbrugge:2017kzv,DiazDorronsoro:2017hti}. Integrals of phase factors such as $e^{iS[g_{\mu\nu}]/\hbar}$ usually do not manifestly converge, but the convergence can be satisfied by shifting the contour of the integral onto the complex plane by applying Picard-Lefschetz theory~\cite{Witten:2010cx}.

According to Cauchy's theorem, if there are no poles within a region on the complex plane, the Lorentzian nature of the integral is preserved even if the integration contour on the complex plane is deformed within such a region. In particular, the path integral can be reformulated to depend solely on the gauge-fixed lapse function $N$ and allows for direct computation.
While the Picard-Lefschetz theory by itself might leave the ambiguity of correct integration paths, utilizing the resurgence theory and Lefschetz thimble analyses in combination can solve this problem~\cite{Honda:2024aro} and enables the precise execution of the gravitational path integral over Lorentzian spacetime. Below, we will briefly explain this Lorentzian framework.

Let us consider a closed, flat, and open Friedmann-Lema\^{i}tre-Robertson-Walker (FLRW) universe with tensor-type metric perturbations whose line element is written as
\begin{equation}\label{eq:metric}
\mathrm{d} s^{2}= -\frac{N^2(t)}{q(t)} \mathrm{d}t^2 
+ q(t)\left[ \Omega_{ij} ({\bf x})+h_{i j} (t \,, {\bf x}) \right] \mathrm{d} x^{i} \mathrm{d} x^{j}\,,
\end{equation}
where $t$ is a time variable, $q(t)=a(t)^2$ is the scale factor squared, $N(t)$ is the lapse function, $\Omega_{ij}$ is the metric of a homogeneous and isotropic 3-dimensional space with curvature constant $K$, $h_{ij}$ represents the tensor perturbation satisfying the transverse and traceless condition, $\Omega ^{i j} h_{i j} = \Omega^{ki}D_k h_{i j} = 0$, $\Omega^{ij}$ is the inverse of $\Omega_{ij}$ and
 $D_i$ is the spatial covariant derivative compatible with $\Omega_{ij}$. 
Given this metric, the gravitational action is expanded up to the second order in the perturbation $h_{i j}$ as
 $S_{\rm GR} = S_{\rm GR}^{(0)} (h^0) +S_{\rm GR}^{(2)} (h^2) +\mathcal{O}(h^3)$:
\begin{align}\displaystyle
S_{\rm GR}^{(0)} &= V_3 \int_{t=t_i}^{t=t_f} \mathrm{d}t
\left(-\frac{3}{4 N} \dot{q}^2+N(3K
-\Lambda q)\right)+S_B \,, 
\label{action_0} \\
S_{\rm GR}^{(2)} &= 
\begin{dcases}
V_3 \int_{t=t_i}^{t=t_f}  N \mathrm{d}t \, \sum_{snlm}\biggl[
\frac{q^2}{8N^2}\left(\dot{h}^{s}_{nlm}\right)^2
-\frac{K}{8}\left((n^2-3)+2 \right)(h^s_{nlm})^2\biggr]\,,   &   K>0  \\
V_3 \int_{t=t_i}^{t=t_f}  N \mathrm{d}t 
\int \frac{\mathrm{d}^3k}{(2\pi)^3} \,\biggl[
\frac{q^2}{8N^2}\left(\dot{h}^{s}_{k}\right)^2
-\frac{1}{8}k^2(h^s_{k})^2\biggr]\,,   &   K=0  \\
V_3 \int_{t=t_i}^{t=t_f}  N \mathrm{d}t \, \sum_{snlm}\biggl[
\frac{q^2}{8N^2}\left(\dot{h}^{s}_{nlm}\right)^2
+\frac{K}{8}\left((n^2+3)-2 \right)(h^s_{nlm})^2\biggr]\,,   &   K<0
\end{dcases}
\end{align}
where $V_3$ is the 3-dimensional volume factor and we add possible boundary contributions $S_B$ localized on the hypersurfaces at $t_{i,f}$. 
For open and closed universes, we have expanded the tensor perturbation $h_{ij}$ in terms of the tensor hyper-spherical harmonics 
with each coefficient $h^s_{nlm}$ being a function of the time $t$, where $s=\pm$ is the polarization label, and the integers ($n$, $l$, $m$) run over the ranges $n\geq3$, $l \in [2,n-1]$, $m \in [-l,l]$ for $K>0$ and the ranges $n\geq 0$, $l\geq 2$, $m \in [-l,l]$ for $K<0$~\cite{Gerlach:1978gy,10.1143/PTP.68.310,Abbott:1986ct}. 
For the flat universe, we have expressed tensor perturbations using eigenfunctions of the flat-space Laplacian and polarization tensors. 
We will restrict our consideration to one mode of tensor perturbations and denote $h^s_{nlm}$ or $h^s_k$ of our interest simply by $h$, suppressing 
the indices $snlm$ or $k$.

Hereafter, we shall construct the gravitational transition amplitude preserving reparametrization invariance through the Batalin-Fradkin-Vilkovisky (BFV) formalism~\cite{Fradkin:1975cq,Batalin:1977pb}.
For the gauge-fixing choice $\dot{N}=0$, 
the BFV path integral reads~\cite{Halliwell:1988wc},
\begin{align}\label{eq:BFV-path-integral}
G[q,h] = \int\! \mathrm{d}N(t_f-t_i) \int 
\mathcal{D}q\mathcal{D}h \exp\left(i S_{{\rm GR}}[N,q,h]
/ \hbar\right)\,,
\end{align}
which is the integral over the proper time $N(t_f-t_i)$ 
between the initial and final configurations. 
The gravitational transition amplitude for the no-boundary and tunneling proposals can be given by Eq.~\eqref{eq:BFV-path-integral} with which the integration is performed from a 3-geometry of zero sizes, i.e. nothing, to a finite one~\cite{Halliwell:1988ik}.

We consider the Dirichlet boundary condition in the Lorentzian path integral, 
%%%%%%%%%%%%%%%%%%%%%%%%%%%%%%
\footnote{In the path integral, it is possible to consider other boundary conditions like Neumann or Robin boundary conditions~\cite{DiTucci:2019dji,DiTucci:2019bui}. However, in this paper, we only consider the simplest Dirichlet boundary conditions, which are suitable for describing quantum cosmogenesis. Moreover, imposing boundary conditions covariantly within the framework of \ac{gr} is not trivial; it is necessary to introduce appropriate boundary terms for the background and perturbations~\cite{York:1972sj,Gibbons:1976ue,Krishnan:2017bte,Brizuela:2023vmb}.}
%%%%%%%%%%%%%%%%%%%%%%%%%%%%%%
where we fix the value of the squared scale factor at the two endpoints,
\begin{equation}\label{eq:Dirichlet-boundary}
q(t_{i}=0)=0, \quad q(t_{f}=1)=q_{f}.
\end{equation} 
In this paper, we focus on the quantum creation of the universe from nothing.
For the action with the above Dirichlet boundary condition, the path integral 
can be exactly evaluated by the time-slicing method. 
On the other hand,
the path integral~\eqref{eq:BFV-path-integral} can be evaluated 
under the semi-classical analysis since the action $S[q,N]$ is quadratic.
We assume the full solution $q(t) = q_s(t) + Q(t)$ where
$Q(t)$ is the Gaussian fluctuation around the semi-classical solution $q_s(t)$.
By substituting it for the action and 
integrating the path integral over $Q(t)$, we can get the explicit expression at the zeroth order of the perturbations $h$~\cite{Feldbrugge:2017kzv}, 
\begin{align}\label{eq:Lorentzian-path-integral}
G^{(0)}[q_{f};0] 
= \int\! \mathrm{d}N \sqrt{\frac{3iV_3}{4\pi\hbar N}}
\exp \left(\frac{i S^{(0)}_\textrm{on-shell}[N]}{\hbar}\right).
\end{align}
where $S_{\rm on-shell}^{(0)}[N]$ is the on-shell action for the background,
\begin{equation}\label{eq:on-shell-action0}
S_\textrm{on-shell}^{(0)}[N]=V_3 \left[ 
\frac{N^3H^4}{4} + N \left( -\frac{3H^2(q_f)}{2} +3K \right) +\frac{1}{N}\left( -\frac{3}{4}(q_f)^2\right) \right]\,,
\end{equation}
with $H^2\equiv \Lambda/3$.
We mainly consider the integration of the lapse function over $N \in (0, \infty)$ and 
this choice of $N$ ensures the causality~\cite{Teitelboim:1983fh}.
On the other hand, 
considering all ranges of the lapse function $N \in (-\infty, \infty)$ is also possible~\cite{DiazDorronsoro:2017hti,Feldbrugge:2017mbc}, but this notorious choice leads to an ambiguity between the no-boundary wave function and the tunneling wave function in the Lorentzian path integral.

We will investigate the behavior of this path integral by using the saddle-point method. 
The derivative of the on-shell action $S_\textrm{on-shell}^{(0)}[N]$
leads to the four saddle points, 
\begin{equation}\label{eq:saddle-points}
N_s=\frac{c_1}{H^2}
\left[\left(-K\right)^{1/2}+c_2\left(q_{f}H^2-K\right)^{1/2}\right],
\end{equation}
with $c_{1,2} \in \{-1 , +1\}$.
Then the saddle-point action $S_\textrm{on-shell}^{(0)}[N_s] $ reads,
\begin{align}\label{eq:saddle-point-action}
S_\textrm{on-shell}^{(0)}[N_s] &=-c_1\frac{2V_3}{H^2}
\left[\left(-K\right)^{3/2} + c_2\left(q_fH^2-K\right)^{3/2} \right].
 \end{align}
In the simplest model to describe the quantum creation of the universe from nothing, 
we usually consider the closed universe and $K>0$.
By using the saddle-point method, for $\textrm{Re}\left[iS_\textrm{on-shell}^{(0)}[N_s]\right]>0$ and $\textrm{Re}\left[iS_\textrm{on-shell}^{(0)}[N_s]\right]<0$, 
we obtain the tunneling and no-boundary wave functions, respectively.
It is found that $c_{1}=+1$ and $c_{1}=-1$
are the tunneling and no-boundary saddle points, respectively.

The saddle points not located on the initial integration contour might still influence the outcome, necessitating further analysis to identify which saddle points are significant. A practical approach involves identifying the steepest descent paths or Lefschetz thimbles associated with the saddle points. By doing so, it's possible to get the integration contour into a combination of thimbles that effectively mirrors the original contour. Notably, the Lefschetz thimble connected to a saddle point, denoted as $\mathcal{J}_{N_s}$, exhibits a critical characteristic: (i) $\textrm{Im}[iS_\textrm{on-shell}^{(0)}[N]] =\textrm{Im}[iS_\textrm{on-shell}^{(0)}[N_s]]$ along $N \in \mathcal{J}_{N_s}$
(ii) $\textrm{Re}[iS_{\rm on-shell}^{(0)}[N]]$ monotonically decreases as we go far away from $N_s$ along $N \in \mathcal{J}_{N_s}$.
Then we rewrite the integral as
\begin{equation}
G^{(0)}[q_{f};0]=\sqrt{\frac{3iV_3}{4\pi \hbar}} \sum_{N_s} n_{N_s} \int_{\mathcal{J}_{N_s}} \frac{\mathrm{d}N}{\sqrt{N}} \exp [iS_\textrm{on-shell}^{(0)}[N]/\hbar] \,, 
\label{eq:Lefschetz-thimbles-decomposition}
\end{equation}
where $n_{N_s}$ is an integer called Stokes multiplier to determine how the saddle $N_s$ contributes.
However, it's not always possible to express the integration contour as a superposition of Lefschetz thimbles, $\mathcal{J}_{N_s}$. In particular, in situations where Stokes phenomena can occur, the decomposition of the Lefschetz thimbles becomes ambiguous.
The Stokes phenomena can occur when there are multiple saddle points with the same imaginary part of the exponent in the integrand, i.e. $\textrm{Im}[iS_\textrm{on-shell}^{(0)}[N_s]]=\textrm{Im}[iS_\textrm{on-shell}^{(0)}[N_s']]$
where $N_s$ and $N_s'$ denote different saddle points. 
In these Stokes phenomena, it is often convenient to deviate slightly from the Stokes lines by adjusting the parameters, and this allows for the observation of changes as one approaches the Stokes lines from various directions.

%%%%%%%%%%%%%%%%%%%%%%%%%%%%%%%%%%%%%%%%%%%%
%%%%%%%%%%%%%%%%%%%%%%%%%%%%%%%%%%%%%%%%%%%%
\begin{figure}[t] 
\centering
\includegraphics[width=0.32\textwidth]{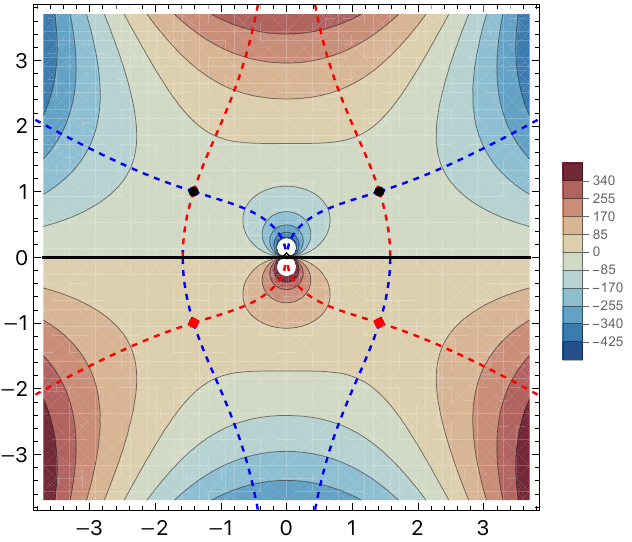}
\includegraphics[width=0.32\textwidth]{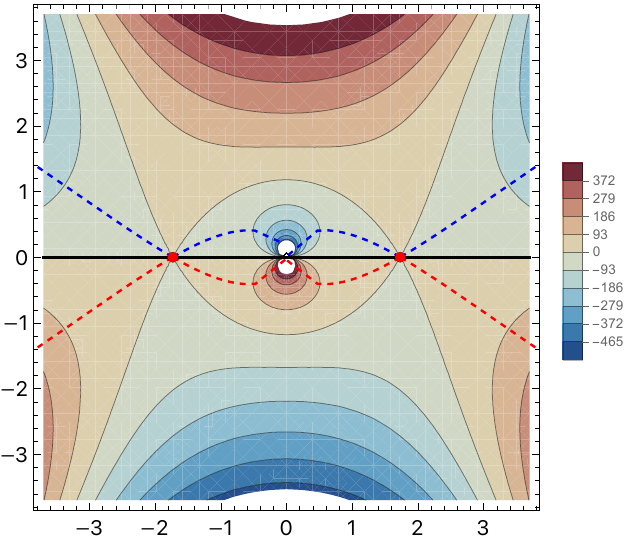}
\includegraphics[width=0.32\textwidth]{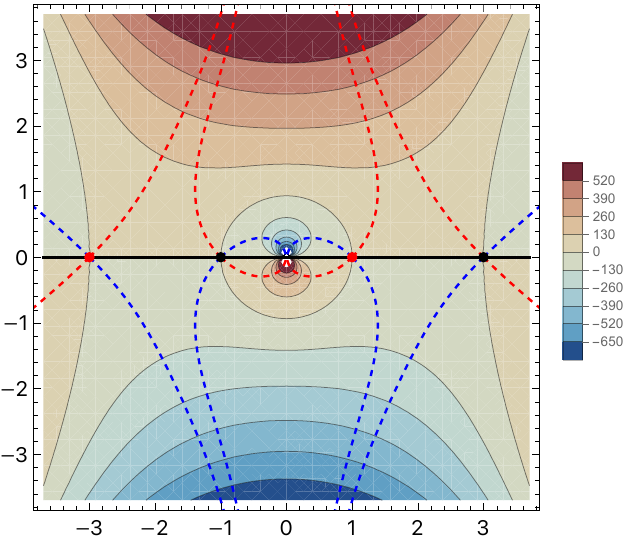}
\caption{
The left contour plot of $\textrm{Re}\left[iS_\textrm{on-shell}^{(0)}[N]\right]$
 over the complex $N$ plane for $\hbar =1 $, $H =1 $, $V_3 =2\pi^2 $, $q_f =3$, and $K =1$. 
A middle and right plot with $K=0$ and $K=-1$ where all other parameters are the same. 
The black (red) circles represent the tunneling (no-boundary) saddle points~\eqref{eq:saddle-points} while the blue (red) lines denote the (dual) Lefschetz thimbles. }
\label{fig:Picard-Lefschetz1}
\end{figure} 
%%%%%%%%%%%%%%%%%%%%%%%%%%%%%%%%%%%%%%%%%%%%
%%%%%%%%%%%%%%%%%%%%%%%%%%%%%%%%%%%%%%%%%%%%

The integration over $N$ in the expression \eqref{eq:Lefschetz-thimbles-decomposition} can be performed by using the Lefschetz thimble analyses. 
We plot $\textrm{Re}\left[iS_\textrm{on-shell}^{(0)}[N]\right]$ for the on-shell action~\eqref{eq:on-shell-action0} over the complex plane in Fig.~\ref{fig:Picard-Lefschetz1}.
The left contour plot corresponds to the no-boundary proposal or tunneling proposal which describes the quantum creation of the closed universe. In this set-up, when we integrate the lapse function over $N \in (0, \infty)$, the Stokes phenomena occur, and the decomposition of the Lefschetz thimbles seems to be ambiguous. However, this ambiguity can be removed by the parameter $\hbar$ to be slightly complex $\hbar=e^{i\theta}$ as usually done in the resurgence theory. The correct Lefschetz thimble $\mathcal{J}_{N_s}$ is understood as the limit $\theta \to\pm\theta$ from nonzero $\theta$ and the contributing saddle points are the same between $\theta >0$ and $\theta <0$~\cite{Honda:2024aro}. 
Thus, the Lefschetz thimble $\mathcal{J}_{N_s}$
passes one complex saddle point with $c_1=c_2=+1$ and leads to the tunneling propagator or tunneling wave function.

In addition, integrating the lapse function over $N \in (-\infty, \infty)$~\cite{DiazDorronsoro:2017hti} in \eqref{eq:Lefschetz-thimbles-decomposition} can provide either the tunneling wave function or the no-boundary wave function as found in Ref.~\cite{Hartle:1983ai} with the Euclidean path integral method, depending on whether the path goes above or below the singularity at $N = 0$. The no-boundary wave function is given by the two complex saddle points with $c_1=-1,\, c_2=-1$ and $c_1=-1,\, c_2=+1$ meaning $\textrm{Im}[N] < 0$. 
After all, the tunneling and no-boundary wave function at the zeroth-order in perturbation are given by~\cite{Feldbrugge:2017kzv,DiazDorronsoro:2017hti},
\begin{align}
G^{(0)}_{\rm T}[q_f] &\simeq
\frac{e^{+i\pi\over4}}{2(q_fH^2 - 1)^{1/4}}
e^{-\frac{4\pi^2}{\hbar H^2} \, -i 4\pi^2 H(q_f - 1/H^2)^{3/2}/\hbar} \quad\quad  (\mbox{tunneling})\label{eq:tunneling-propagator}\,, \\
G^{(0)}_{\rm NB}[q_f] &\simeq
\frac{e^{+\frac{4\pi^2}{\hbar H^2}} }{(q_f H^2 - 1)^{1/4}}\cos\left[ \frac{4\pi^2}{\hbar H^2} \left(q_fH^2 - 1\right)^{3/2} + \frac{3 \pi}{4} \right] \quad (\mbox{no-boundary})\label{eq:no-boundary-propagator}\,,
\end{align}
where the first equation considered the integral over $N \in (0, \infty)$, while the latter considered the entire integral range, and we set $V_3=2\pi^2$, $K=1$ and $q_f > 1/H^2$.

On the other hand, the middle and right plots in Fig.~\ref{fig:Picard-Lefschetz1} correspond to the flat and open universe, and the corresponding saddle points are real, and consequently the typical behaviour of these wave functions shows significant differences from the traditional scenario of quantum cosmology. 
The no-boundary and tunneling wave functions behave exponentially as a function of the cosmological constant $H^2\equiv \Lambda/3$, whereas the open or flat universe wave function behaves oscillatively as a function of the cosmological constant~\cite{Coule:1999wg,Linde:2004nz}. 
In the saddle-point approximation, we have the following wave function, 
\begin{align}
G^{(0)}_\textrm{open}[q_f] &\simeq
\sqrt{\frac{3\pi i}{2\hbar}\frac{H^2}{\left(
q_{f}H^2+1\right)^{1/2}\pm 1}}\exp \left[-\frac{4\pi^2i}{\hbar H^2} \left(\left(q_fH^2 + 1\right)^{3/2} \pm 1\right)\right] \quad (\mbox{open})\label{eq:wave-function-open}\,,\\ 
G^{(0)}_\textrm{flat}[q_f] &\simeq \sqrt{\frac{3\pi iH}{2q_f^{1/2}\hbar}}
e^{-\frac{i}{\hbar}\cdot 4\pi^2q_f^{3/2}H} \quad\quad  (\mbox{flat})\label{eq:wave-function-flat}\,,
\end{align}
where we assumed that the lapse integration is $N \in (0, \infty)$.
In the next section, we will consider perturbation problems for these cosmological wave functions.

%%%%%%%%%%%%%%%%%%%%%%%%%%%%%%%%%%%%
%%%%%%%%%%%%%%%%%%%%%%%%%%%%%%%%%%%
\section{Perturbation crisis in quantum cosmology }
\label{sec:Perturbation-analysis}
%%%%%%%%%%%%%%%%%%%%%%%%%%%%%%%%%%%%
%%%%%%%%%%%%%%%%%%%%%%%%%%%%%%%%%%%%
In the previous section, we have discussed the Lorentzian formulations of the wave function
of the closed, flat, and open universe at the zeroth order of perturbations
in path integrals of quantum gravity. 
Hereafter, we discuss such wave functions $\Psi[g]$, including tensor perturbations. 
In GR~\cite{Feldbrugge:2017kzv,DiazDorronsoro:2017hti}, it has been shown that for the wave functions of the closed universe, the linearized perturbations around the background 
are governed by an inverse Gaussian distribution, which leads to divergent correlation functions, and thus the perturbation is uncontrollable. 
In this section, we demonstrate that the inverse Gaussian wave function for linearized tensor perturbations in the Lorentzian path integral is inevitable, as long as the regularity of the on-shell gravitational action is required in quantum cosmology.

We perform the Lorentzian path integral for Eq.~\eqref{eq:BFV-path-integral} up to the second
order in the tensor perturbation where we neglect the back-reaction of the linearized tensor 
perturbations. 
First, as previously discussed, we integrate the path integral with respect to the background $q(t)$ and the tensor perturbation $h(t)$ around each classical solution, and evaluate the total on-shell action including the background and the tensor perturbation as
\begin{align}\label{eq:on-shell-action}
\begin{split}
S_\textrm{on-shell}[q,h,N] &=
S_\textrm{on-shell}^{(0)}[q,N]
+S_\textrm{on-shell}^{(2)}[h,N] +\mathcal{O}(h^3) \,.
\end{split} 
\end{align}
Next, we integrate $\exp \left({{iS_\textrm{on-shell}[q,h,N] }/{\hbar}}\right)$ over the lapse function $N$. In this step, we utilize the Picard-Lefschetz method, which complexifies $N$ and selects complex integration contours along the Lefschetz thimbles $\mathcal{J}_{N_s}$. Since $\mathcal{J}_{N_s}$ ensures the convergence of the integral, 
we can efficiently perform the Lorentzian path integral at the perturbative level.
In particular, assuming that the perturbations do not affect the background spacetime, the saddle points of the total on-shell action can be approximated nearly as $N_s$ \eqref{eq:saddle-points} of the background. Since it is confirmed numerically in the full analysis, we can approximately use $N_s$ \eqref{eq:saddle-points} of the background.

We write down the second-order action for the tensor perturbation,
\begin{align}
\begin{split}\label{action2}
S_{\rm GR}^{(2)}[h,N] =  V_3 \int_{0}^{1} N \mathrm{d}t \, 
\left\{ \frac{q^2}{8N^2}\dot{h}^2
-\frac{\alpha_{\rm mode}}{8}h^2\right\} \,,
\end{split}
\end{align}
where $\alpha_{\rm mode} \in \{K\left((n^2-3)+2\right), k^2, -K\left((n^2+3)-2\right)\}$ for the closed, 
flat and open universe, respectively.
We can evaluate the on-shell action $S_\textrm{on-shell}^{(2)}[N] $,
\begin{align} \label{eqn:S2onshell}
S_\textrm{on-shell}^{(2)}[N] &=
  \frac{\pi^2}{4} \left[ q^2 \frac{h \dot{h}}{N} \right]^1_0\,,
  \end{align}
where we set $V_3=2\pi^2$, and  performed the integration by parts for the action~\eqref{action2}
and used the equation of motion for $h(t)$.
For convenience, let us introduce $\chi(t)=q(t) h(t) $ and
write the equation of motion for $\chi(t)$ as
\begin{align}\label{eq:eom-original}
\frac{\ddot{\chi}}{N^2}  + \left[ \frac{\alpha_{\rm mode}}{q^2} - \frac{1}{N^2} \frac{\ddot{q}}{q}\right]\chi= 0 \,.
\end{align} 
Given the classical solution for the background $q(t)=N^2H^2 t(t-1)+q_ft$
which satisfies the boundary condition $q(0)=0$
and $q(1)=q_f$, we have the solution for 
the above equation (\ref{eq:eom-original}) as~\cite{Matsui:2022lfj},
\begin{align}\label{eq:gr-solution}
&\chi(t)=\sqrt{ \left(H^2 N^2 t (t -1)+q_ft \right) 
\left(H^4 N^2t(t -1)  +H^2 q_ft+ \alpha_{\rm mode} \right)}\\
&\times  \Biggl\{ C_1 \left(\frac{H^2 N^2 (t -1)+q_f}{t}\right)^{\delta \over2} \sqrt{\frac{H^2 N^2 (2t -1)+\sqrt{H^4 N^4+N^2 \left(-4 \alpha_{\rm mode} -2 H^2 q_f\right)+q_f^2}+q_f}{H^2 N^2 (2t -1)-\sqrt{H^4 N^4+N^2 \left(-4 \alpha_{\rm mode} -2 H^2 q_f\right)+q_f^2}+q_f}}\notag \\
&+C_2 \left(\frac{t}{H^2 N^2 (t-1)+q_f}\right)^{
\delta \over2}\sqrt{\frac{H^2 N^2 (2t -1)-\sqrt{H^4 N^4+N^2 \left(-4 \alpha_{\rm mode} -2 H^2 q_f\right)+q_f^2}+q_f}{H^2 N^2 (2t -1)+\sqrt{H^4 N^4+N^2 \left(-4 \alpha_{\rm mode} -2 H^2 q_f\right)+q_f^2}+q_f}}\Biggr\}\notag ,
\end{align}
where $C_{1,2}$ are constants, and we have defined,
\begin{align}
\delta[N]=\frac{\sqrt{\left(H^2 N^2-q_f\right)^2-4N^2\alpha_{\rm mode}}}
{\left(H^2 N^2-q_f\right)}\notag =-\sqrt{1-\frac{4N^2\alpha_{\rm mode}}
{\left(q_f-N^2H^2\right)^2}}\,,  
\end{align}
where $\textrm{Re}[\delta[N]]<0$ is negative for all
complex $N$ plane away from the branch cuts~\cite{Feldbrugge:2017mbc}.

When we substitute the saddle-points $N_s$ \eqref{eq:saddle-points} into $\delta[N]$, we have 
\begin{align}\displaystyle
\delta[N_s]=-\sqrt{\frac{\alpha_{\rm mode} +K}{K}}
\end{align}
For the closed and flat universe, $\delta[N_s]$ is negatively real whereas for the open universe, $\delta[N_s]$ 
is imaginary. Hereafter, 
we simply drop the dependence of $N$ in $\delta[N]$ and write $\delta$.

To estimate the on-shell action, we only need the values of $\chi(t)$ and $\dot{\chi}(t)$ at $t = 0, 1$. 
The on-shell action for the solution~\eqref{eq:gr-solution} is written as
\begin{align}\label{eq:semiclassical-action}
&S_\textrm{on-shell}^{(2)}[N]=-\frac{\pi^2 \alpha_{\rm mode} }{8 N}\Biggl[
C_1^2 q_f^{\delta} \left(\sqrt{\left(q_f-H^2 N^2\right)^2-4 \alpha_{\rm mode}  N^2}+H^2 N^2+q_f\right)+C_2^2q_f^{-\delta}  \\
&\times \left(-\sqrt{\left(q_f-H^2 N^2\right)^2-4 \alpha_{\rm mode}  N^2}+H^2 N^2+q_f\right) +2 C_1C_2 \left(H^2 N^2+q_f\right)\Biggr]
-{\pi^2\over 4 N}q^2\dot{h}h \Bigr|_{t=0}\notag\,.
\end{align}
Near the boundary $t = 0$, the solution~\eqref{eq:gr-solution} behaves as 
\begin{equation}\label{eq:gr-solution0}
\chi(t)\propto C_1\, F_1[N] t^{\frac{1}{2}(1-\delta)}
+C_2\, F_2[N]t^{\frac{1}{2}(1+\delta)}\quad (t \to 0)\, ,
\end{equation}
where $F_1[N]$, $F_2[N]$ are functions of $N$ whose explicit form can be derived from the general solution~\eqref{eq:gr-solution}. From this expression one can show that the contribution of $t=0$ to the on-shell action (\ref{eqn:S2onshell}) contains terms of the form $\propto C_1t^{-\delta}$, $\propto C_1C_2$ and $\propto C_2^2t^{\delta}$, and thus is finite if and only if $C_1=0$ or $C_2=0$ for $\textrm{Re}[\delta]>0$ or for $\textrm{Re}[\delta]<0$, respectively. 
As we mentioned $\delta[N_s]$ is real and negative for the closed and flat universe so that 
we should take $C_2=0$. On the other hand, for the open universe, $\delta[N_s]$ is imaginary and at the saddle points $N_s$ \eqref{eq:saddle-points}, to take $C_1=0$ seems to be possible. We shall discuss this point in more detail later.

Setting either $C_1=0$ (for $\textrm{Re}[\delta]>0$) or $C_2=0$ (for $\textrm{Re}[\delta]<0$), and then fixing the remaining integration constant $C_2$ or $C_1$, respectively, by $\chi(1)=q_fh_f$ for the solution~\eqref{eq:gr-solution}, we obtain the on-shell action. 
For convenience, by using $\delta[N]$ we rewrite the on-shell action as follows, 
\begin{align}\label{eq:gr-perturbation-onshell-action}
S_\textrm{on-shell}^{(2)}[N]=\begin{dcases} 
-\frac{\pi ^2 q_fh_f^2\alpha_{\rm mode} \left(\left(H^2 N^2-q_f\right)\delta[N]
+H^2 N^2+q_f\right)}
{8 N \left(\alpha_{\rm mode} +H^2 q_f\right)}\quad  
\textrm{for}\ \textrm{Re}[\delta]>0 \\ 
-\frac{\pi ^2 q_fh_f^2\alpha_{\rm mode} \left(-\left(H^2 N^2-q_f\right)\delta[N]+H^2 N^2+q_f\right)}{8 N \left(\alpha_{\rm mode} +H^2 q_f\right)}\quad  \textrm{for}\ \textrm{Re}[\delta]<0\,, \end{dcases}
\end{align}
where we include the $N$ dependence of $\delta$ in the on-shell action and write $\delta[N]$ again.

%%%%%%%%%%%%%%%%%%%%%%%%%%%%%%%%%%%%%%%%%%%%%%%%%%%%%%%%%%%
%%%%%%%%%%%%%%%%%%%%%%%%%%%%%%%%%%%%%%%%%%%%%%%%%%%%%%%%%%%
\subsection{No-boundary and tunneling proposals}
%%%%%%%%%%%%%%%%%%%%%%%%%%%%%%%%%%%%%%%%%%%%%%%%%%%%%%%%%%%
%%%%%%%%%%%%%%%%%%%%%%%%%%%%%%%%%%%%%%%%%%%%%%%%%%%%%%%%%%%

Hereafter, we shall concentrate on the no-boundary and tunneling proposals where we consider the quantum creation of the universe from nothing, i.e. $q(t_{i}=0)=0$ and $K=1$. 
In particular, we do not take into account a possible shift of the position of the saddle points in the complex-$N$ plane due to the tensor perturbations and simply evaluate $S_{\rm on-shell}^{(2)}[N]$ for $\chi(\tau)$ at the background saddle points $N_s$ \eqref{eq:saddle-points} in the complex-$N$ plane. As previously commented, this would be a good approximation as long as the back-reaction of the perturbations is negligible.

Now, we simply set $K=1$ and consider the integration of the lapse function over $N \in (0, \infty)$, where the Lefschetz thimble $\mathcal{J}_{N_s}$ passes one complex saddle-point $N_s$ \eqref{eq:saddle-points} with $c_1=c_2=+1$ (see left plot in Fig.~\ref{fig:Picard-Lefschetz1}), 
\begin{align}\label{eq:tunneling-saddle}
N_\textrm{T}=\frac{1}{H^2}
\left[i+\left(q_{f}H^2-1\right)^{1/2}\right], 
\end{align}
and this leads to the tunneling propagator or tunneling wave function.
Taking the tunneling saddle point 
$N_\textrm{T}$~\eqref{eq:tunneling-saddle} for 
the on-shell action of the perturbations~\eqref{eq:gr-perturbation-onshell-action} with $C_2=0$,
we have
\begin{align}\label{eq:gr-perturbation-onshell-action-ts}
\frac{i}{\hbar} S_\textrm{on-shell}^{(2)}[N_\textrm{T}]
&=-\frac{i\pi^2 q_fh_f^2\alpha_{\rm mode} \left(-i\delta[N]+\sqrt{q_fH^2 -1}\right)}{4 \hbar  \left(q_fH^2+\alpha_{\rm mode}\right)} \notag\\
&=+\frac{\pi^2 q_fh_f^2\alpha_{\rm mode} \left(\sqrt{\alpha_{\rm mode}+1}-i \sqrt{q_fH^2 -1}\right)}{4 \hbar  \left(q_fH^2+\alpha_{\rm mode}\right)} \notag\\
&\approx \frac{\pi^2n(n^2-1)}{4\hbar H^2}\left[ 1 - in^{-1}q_f^{1/2}H + \cdots \right]h_f^2\,,
\end{align}
where we took the super-horizon mode ($n\ll q_f^{1/2}H$)
and $q_f\gg 1/H^2$ in the last expression. 
The real part of \eqref{eq:gr-perturbation-onshell-action-ts} is positive and thus induces an inverse Gaussian distribution for the tensor perturbations. In contrast, at the no-boundary saddle point $N_\textrm{HH}$ the tensor perturbations follow a Gaussian distribution according to the on-shell action~\eqref{eq:gr-perturbation-onshell-action-ts}. Although the contribution of the no-boundary saddle points to the correlation functions of the linearised perturbations is finite, we must consider the lapse $N$ integration over $N \in (-\infty, \infty)$~\cite{DiazDorronsoro:2017hti} and pass through the tunneling saddle point $N_\textrm{T}$ and the branch cuts, which again lead to divergent correlation functions of linearised perturbations~\cite{Feldbrugge:2017mbc}. As a result, the inclusion of tensor perturbations in the no-boundary proposal~\eqref{eq:no-boundary-propagator} leads to inconsistencies. Attempts to solve these problems by modifying the boundary conditions of the 
background and fixing the imaginary initial momentum~\cite{DiTucci:2019dji,DiTucci:2019bui} depart from the original framework describing quantum creation from nothing ($q(t_{i}=0)=0$) and do not save the tunneling proposal. This is a brief overview of the dilemmas that the no-boundary and tunneling proposals face due to linearised perturbations in the Lorentzian quantum gravity.

%%%%%%%%%%%%%%%%%%%%%%%%%%%%%%%%%%%%%%%%%%%%%%%%%%%%%%%%%%%
%%%%%%%%%%%%%%%%%%%%%%%%%%%%%%%%%%%%%%%%%%%%%%%%%%%%%%%%%%%
\subsection{Open and flat universe}
%%%%%%%%%%%%%%%%%%%%%%%%%%%%%%%%%%%%%%%%%%%%%%%%%%%%%%%%%%%
%%%%%%%%%%%%%%%%%%%%%%%%%%%%%%%%%%%%%%%%%%%%%%%%%%%%%%%%%%%

Next, we consider the open and flat universe. We study the quantum creation of the universe from nothing where spatial sections are compact but the curvature is characterized as flat or open ($K=-1,0$)~\cite{Coule:1999wg,Linde:2004nz}. In such scenarios of quantum cosmology, the saddle points $N_s$ \eqref{eq:saddle-points} of the on-shell action $S_\textrm{on-shell}^{(0)}[N]$ are deformed to be real. As the previous discussion, we have shown that complex saddle points where $\textrm{Im}[N]>0$ pose challenges in the context of perturbations so one might think that real saddle points may offer solutions to the perturbation problems.
However, we will show that the real saddle points do not fully solve the perturbation problems.

First, we consider the open universe and simply set $K=-1$.
When we assume the integration of the lapse function over $N \in (0, \infty)$, the Lefschetz 
thimble $\mathcal{J}_{N_s}$ passes two real saddle-points (see right plot in Fig.~\ref{fig:Picard-Lefschetz1}), 
\begin{equation}\label{eq:open-saddle}
N_\textrm{open}=\frac{1}{H^2}
\left[c_3+\left(q_{f}H^2+1\right)^{1/2}\right],
\end{equation}
with $c_{3} \in \{-1 , +1\}$.
Taking the above saddle point for the on-shell action~\eqref{eq:gr-perturbation-onshell-action} with $C_2=0$, we have
\begin{align}\label{eq:gr-perturbation-onshell-action-os}
\frac{i}{\hbar} S_\textrm{on-shell}^{(2)}[N_\textrm{open}] 
&=\frac{\pi^2q_fh_f^2\alpha_{\rm mode}\left((\delta[N] -1)\left(c_3 \sqrt{q_fH^2 +1}+1\right)-q_fH^2\right)}{4 \left(c_3+\sqrt{q_fH^2+1}\right) \left(q_fH^2+\alpha_{\rm mode}\right)}\notag\\
&=-c_3\frac{\pi^2q_fh_f^2\alpha_{\rm mode}
\left(\frac{iq_fH^2 }{1+c_3\sqrt{q_fH^2 +1}}+i \sqrt{1-\alpha_{\rm mode}}+i\right)}
{4\hbar\left(q_fH^2 +\alpha_{\rm mode}\right)}\,.
\end{align}
The real part of (\ref{eq:gr-perturbation-onshell-action-os}) becomes positive or negative depending on the sign of $c_3$. For $c_3=+1$ meaning the tunneling saddle point, the
tensor perturbations exhibit an inverse Gaussian distribution. 
In this sense, the quantum creation of the open universe has a perturbative problem and is naively not consistent with cosmological observations.

Next, let us consider the flat universe and take $K=0$.
When we assume the integration of the lapse function over $N \in (0, \infty)$, the Lefschetz 
thimble $\mathcal{J}_{N_s}$ passes one real saddle-point (see middle plot in Fig.~\ref{fig:Picard-Lefschetz1}), 
\begin{equation}\label{eq:flat-saddle}
N_\textrm{flat}^+=+\frac{q_{f}^{1/2}}{H}.
\end{equation}
Taking the above saddle point for the on-shell action~\eqref{eq:gr-perturbation-onshell-action} with $C_2=0$, we have
\begin{align}\label{eq:gr-perturbation-onshell-action-fs}
\frac{i}{\hbar} S_\textrm{on-shell}^{(2)}[N_\textrm{flat}] 
&=+\frac{\pi ^2q_fh_f^2\alpha_{\rm mode} \left(-\sqrt{\alpha_{\rm mode} }-i\sqrt{q_f}H\right)}
{4\hbar \left(\alpha_{\rm mode}+q_fH^2\right)} \notag\\
&\approx -\frac{\pi^2k^{3}}{4\hbar H^2}\left[ 1 + i q_f^{1/2}Hk^{-1} + \cdots \right] h_f^2\,,
\end{align}
where we took the super-horizon mode ($k\ll q_f^{1/2}H$)
and $q_f\gg 1/H^2$ in the last expression. 
Unlike the closed universe, the real component of Eq.~(\ref{eq:gr-perturbation-onshell-action-fs}) is negative, which results in the tensor perturbations following a Gaussian distribution and indicates that the perturbations can be suppressed.
This result corresponds to the standard analysis of the inflation theory taking the Bunch-Davies vacuum. 
However, beyond the saddle-point approximation, we must integrate the lapse function $N$ over all the complex planes, and the perturbation problem appears again.
For instance, considering the variation around the saddle point $N_\textrm{flat}$ as, 
\begin{equation}
N_\textrm{flat}=\frac{q_{f}^{1/2}}{H}+i\frac{\Delta_N}{H^2}, 
\end{equation}
where $i{\Delta_N}/{H^2}$ is the complex variance of the lapse function.
If we set $\Delta_N=1$, we obtain the tunneling saddle point $N_\textrm{T}$~\eqref{eq:tunneling-saddle} and the tensor perturbation exhibits an inverse Gaussian distribution. Only at the saddle point, the stability of the perturbations is ensured. Furthermore, by using the on-shell action~\eqref{eq:gr-perturbation-onshell-action}, it is easy to show that the perturbation is again inverse Gaussian if the quantum creation of the flat universe is taken as the zero limit $K\to 0^{+}$ from which the curvature is positive~\cite{DiTucci:2019xcr}.

Furthermore, when we take the integration of the lapse function over $N \in (-\infty, \infty)$, the Lefschetz 
thimble $\mathcal{J}_{N_s}$ passes two real saddle-points, 
\begin{equation}\label{eq:flat-saddles}
N_\textrm{flat}^{\pm}=\pm \frac{q_{f}^{1/2}}{H}.
\end{equation}
Taking the above saddle points for the on-shell action~\eqref{eq:gr-perturbation-onshell-action} with $C_2=0$, and assuming the super-horizon mode ($k\ll q_f^{1/2}H$)
and $q_f\gg 1/H^2$, we obtain the following expression, 
\begin{align}\label{eq:gr-perturbation-onshell-action-fs-two}
\frac{i}{\hbar} S_\textrm{on-shell}^{(2)}[N_\textrm{flat}^{\pm}] 
&=\pm \frac{\pi ^2q_fh_f^2\alpha_{\rm mode} \left(-\sqrt{\alpha_{\rm mode} }-i\sqrt{q_f}H\right)}
{4\hbar \left(\alpha_{\rm mode}+q_fH^2\right)} \notag\\
&\approx \mp \frac{\pi^2k^{3}}{4\hbar H^2}\left[ 1 + i q_f^{1/2}Hk^{-1} + \cdots \right] h_f^2\,.
\end{align}
For the saddle point $N_\textrm{flat}^{-}$, 
the real component of Eq.~(\ref{eq:gr-perturbation-onshell-action-fs-two}) is positive, 
which results in the tensor perturbations following an inverse Gaussian distribution once more.

Finally, we will briefly discuss the primordial inflation case. 
For simplicity, we assume the flat spatial curvature $K=0$ and 
the finite value of the initial scale factor $q(t_{i}=0)=q_i$.
When we assume the integration of the lapse function over $N \in (0, \infty)$, the Lefschetz 
thimble $\mathcal{J}_{N_s}$ passes two real saddle-points~\cite{Feldbrugge:2017kzv}, 
\begin{equation}\label{eq:inf-saddle}
N_\textrm{inf}=\frac{1}{H}
\left[c_3q_{i}^{1/2}+q_{f}^{1/2}\right],
\end{equation}
with $c_{3} \in \{-1 , +1\}$. Similar to the open universe case, the perturbations are found to be stable at one saddle point and unstable at another~\cite{DiTucci:2019xcr}.
One difference is that the regularity of the on-shell action does not have to be strictly enforced. Indeed, since it does not take $q_i=0$, we may take the divergent solutions, but
it makes a large contribution to the on-shell action. Therefore, the above conclusions would still hold when the initial size of spacetime is small, $q_i\ll1 $.
See the Appendix~\ref{appendix:Bunch-Davies} for a comparison with standard perturbation analysis of inflation theory.
We have seen that the quantum creation of closed, flat, and open universes from nothing, and the beginning of primordial inflation have serious perturbation problems.
In the next section, we will discuss the perturbation problem beyond \ac{gr}.

%%%%%%%%%%%%%%%%%%%%%%%%%%%%%%%%%%%%
%%%%%%%%%%%%%%%%%%%%%%%%%%%%%%%%%%%%
\section{Trans-Planckian physics and Modified dispersion relations}
\label{sec:TPP-modified}
%%%%%%%%%%%%%%%%%%%%%%%%%%%%%%%%%%%%
%%%%%%%%%%%%%%%%%%%%%%%%%%%%%%%%%%%%

In this section, we will discuss the perturbation problem in 
Lorentzian quantum cosmology by assuming modified dispersion relations based on the trans-Planckian physics~\cite{Martin:2000xs,Brandenberger:2000wr,
Niemeyer:2000eh,Martin:2002kt,Ashoorioon:2004vm,Ashoorioon:2011eg}. 
In our previous work~\cite{Matsui:2022lfj}, we demonstrated that the perturbation problem of the no-boundary and tunneling proposal in Lorentzian quantum cosmology is hard to overcome with the trans-Planckian physics modifying the dispersion relation at wavelengths smaller than the Planck scale.
In this section, we extend the previous analysis to the general case where we consider the quantum creation of the closed, flat, and open universe and the primordial inflation.

If new physics appears at short distances, the dispersion relation of the perturbation would be modified and we can assume that such modified dispersion relation takes the form $\omega^2 = {\cal F}\, ({k_{\rm phys}})$~\cite{Brandenberger:2012aj}, where $k_{\rm phys}=\alpha_{\rm mode} ^{1/2}/{q^{1/2}}$ is the physical wavenumber. Since the physical momentum diverges at $q \to 0$, the dispersion relation would drastically change. For instance, we can introduce the modified dispersion relation, including the trans-Planckian cutoff~\cite{Niemeyer:2000eh}, 
\begin{align}\label{Modified-dispersion-cutoff}
{\cal F}\, (k_{\rm phys})=\begin{cases} 
k_{\rm phys}^2 \quad \textrm{for}\quad k_{\rm phys}^2\ll {\cal M}_{\rm UV}^2 \\ 
{\cal M}_{\rm UV}^2\quad \textrm{for}\quad k_{\rm phys}^2\gg {\cal M}_{\rm UV}^2, \end{cases}
\end{align}
where ${\cal M}_{\rm UV}$ is the trans-Planckian or \ac{uv} cutoff scale. 
In the limit $q\to 0$, correspondingly $k_{\rm phys} \to \infty$, the dispersion relation for all modes is modified from that in \ac{gr}. 
A concrete example of such modified dispersion relations including the trans-Planckian cutoff is the Unruh dispersion relation~\cite{Unruh:1994je}. Another well-known modified dispersion relation is the generalized Corley-Jacobson dispersion relation~\cite{Martin:2000xs}, 
which takes the following form, 
\begin{align}\label{Modified-dispersion-general}
{\cal F}\, (k_{\rm phys})
= k_{\rm phys}^2
+k_{\rm phys}^2\sum^{p}_{j=1}b_j\left(\frac{k_{\rm phys}^2}{{\cal M}_{\rm UV}^2}\right)^{j}\,,
\end{align}
where the right-hand side should be non-negative for all $k_{\rm phys}^2 \geq 0$ to avoid instability. 
%%%%%%%%%%%%%%%%%%%%%%%%%%%%%%%%%
\footnote{
We have assumed that $b_j>0$ with $j=1...p$ to avoid instability, but if we could take $b_p<0$ the perturbation problems in quantum cosmology would be solved since the regularity of the on-shell action is improved. However, it is difficult to make such an assumption in the well-defined theories of gravity. }
%%%%%%%%%%%%%%%%%%%%%%%%%%%%%%%%%
This was introduced in the context of black hole 
physics~\cite{Corley:1996ar,Corley:1997pr}. 
The modified dispersion relation~\eqref{Modified-dispersion-general} can be introduced by higher-dimensional operators in higher-curvature theories of gravity such as \ac{hl} gravity theory~\cite{Horava:2009uw}. Therefore, our analysis is expected to be applicable to the analysis of the modified gravity theories.

We will consider the generalized Corley-Jacobson dispersion relation~\cite{Corley:1996ar,Corley:1997pr,Martin:2000xs} as examples of the modified dispersion relations,
and use the second-order action $S_{\rm GR}^{(2)}[h,N] $ for the tensor perturbation $h$ with the dispersion relation $\omega^2 = {\cal F}\, (k_{\rm phys})$ in~\eqref{Modified-dispersion-general}. To simplify our analysis, we assume that the dispersion relation~\eqref{Modified-dispersion-general} only contains the last term of the sum and consider $p=2$. Thus, we have the following equation of motion,
\begin{align}\label{eq:eom-jc}
\frac{\ddot{\chi}}{N^2}  + \left\{ \frac{\alpha_{\rm mode}}{q^2}
\left[1+b_2\left(\frac{\alpha_{\rm mode}}{q
{\cal M}_{\rm UV}^2}\right)^{2}\, \right]- \frac{1}{N^2} \frac{\ddot{q}}{q}\right\}\chi= 0 \,.
\end{align}

Now, we consider the \ac{uv} regime $\alpha_{\rm mode}/q \gg {\cal M}_{\rm UV}^2$, and simply seek the solutions for the tensor perturbations and calculate the contribution of the \ac{uv} boundary ($t=0$), to the on-shell action. 
We will consider the following equation of motion in the \ac{uv} regime,
\begin{align}
\frac{\ddot{\chi}}{N^2}  + \left\{ \frac{\alpha_{\rm mode}}{q^4}\beta
- \frac{1}{N^2} \frac{\ddot{q}}{q}\right\}\chi= 0 \,,
\end{align}
where we set $\beta=b_2\alpha_{\rm mode}^2/{\cal M}_{\rm UV}^4$.
By using the background solution $q(t)=N^2H^2 t(t-1)+q_ft$, 
the perturbative solution reads  
\begin{align}\label{eq:gcj-solution}
\chi(t)&=C_3\left(N^{2} H^{2} \left(t-1\right)+q_f\right)^{\zeta_1}t^{\, \zeta_2} 
\exp \left[{-\frac{\sqrt{-\alpha_{\rm mode}\beta}\, 
N \left(N^{2}H^{2}\left(2t-1\right) +q_f\right)}
{\left(N^{2} H^{2}-q_f\right)^{2}\left(N^{2}H^{2}
\left(t-1\right)+q_f\right)t}}\right] \notag\\
&+C_4 \left(N^{2} H^{2} \left(t-1\right)+q_f\right)^{\zeta_2}\tau^{\, \zeta_1}
\exp \left[{+\frac{\sqrt{-\alpha_{\rm mode}\beta}\, N \left(N^{2}H^{2} \left(2t-1\right) +q_f\right)}{\left(N^{2} H^{2}-q_f\right)^{2}\left(N^{2} H^{2}\left(t-1\right)+q_f\right)t}}\right], 
\end{align}
where $C_{3,4}$ are constants, and we define $\zeta_{1,2}$ as, 
\begin{align}
\zeta_1=1 - 2 {\frac{N^{3} H^{2} \sqrt{-\alpha_{\rm mode}\beta}}{\left(N^{2} H^{2}-q_f\right)^{3}}}, \quad 
\zeta_2=1 + 2 \frac{N^{3} H^{2} \sqrt{-\alpha_{\rm mode}\beta}}{\left(N^{2} H^{2}-q_f\right)^{3}}.
\end{align}

As previously shown in Section~\ref{sec:Perturbation-analysis}, the on-shell action has two contributions, one from the \ac{uv} ($t=0$) and the other from the \ac{ir} ($t=1$). In order to estimate the \ac{uv} contribution, we approximate the above solution~\eqref{eq:gcj-solution} near $t = 0$, 
\begin{equation}\label{eq:gcj-solution0}
\chi(t)\propto C_3\, F_3[t,N] e^{-{\lambda\over t}}
+C_4\, F_4[t,N]e^{+{\lambda\over t}}\, ,
\end{equation}
where $F_3[t,N]$ and $F_4[t,N]$ are polynomial functions of $t$ whose coefficients depend on $N$, and $\lambda=\sqrt{-\alpha_{\rm mode}\beta}N/(H^2N^2-q_f)^{2}$. It is clear that the \ac{uv} ($t=0$) contribution to the on-shell action with $C_3= 0$ vanishes for $\textrm{Re}[\lambda]<0$ whereas that with $C_4= 0$ vanishes for $\textrm{Re}[\lambda]>0$. Other choices lead to a divergent on-shell action. Hereafter, we adopt the choices that avoid a divergent on-shell action ($C_3= 0$ for $\textrm{Re}[\lambda]<0$ or $C_4= 0$ for $\textrm{Re}[\lambda]>0$) and, as a result, the \ac{uv} ($t=0$) contribution is zero.

For modes satisfying $\beta\gg q_f^2$ ($\alpha_{\rm mode}/q_f\gg {\cal M}_{\rm UV}^2$ for $b_2=\mathcal{O}(1)$), we can evaluate not only the \ac{uv} ($t=0$) contribution but also the \ac{ir} ($t=1$) one to the on-shell action by using the solution~\eqref{eq:gcj-solution}.
As explained above, to avoid divergences of the on-shell action, we have supposed that $C_3=0$ for $\textrm{Re}[\lambda]<0$ and that $C_4=0$ for $\textrm{Re}[\lambda]>0$. By imposing $\chi(1)=q_fh_f$ to normalize the overall factor for the solution~\eqref{eq:gcj-solution} with $C_3=0$ or $C_4=0$, we obtain
\begin{align}\label{eq:gcj-semiclassical-action}
S_\textrm{on-shell}^{(2)}[N]=\begin{cases} 
-\frac{\pi ^2 }{4}\sqrt{-\alpha_{\rm mode} \beta }h_f^2
\quad
\textrm{for}\ \ \textrm{Re}[\lambda]<0 \\ 
+\frac{\pi ^2 }{4}\sqrt{-\alpha_{\rm mode} \beta }h_f^2
\quad \textrm{for}\ \ \textrm{Re}[\lambda]>0, \end{cases}
\end{align}
Indeed, we obtain inverse Gaussian or Gaussian distribution for the tensor perturbations as
\begin{align} \label{eqn:p=2inverseGorG}
\frac{i}{\hbar} S_\textrm{on-shell}^{(2)}[N]=\begin{cases} 
+\frac{\pi ^2 }{4\hbar}{\alpha_{\rm mode}^{3/2}b_2^{1/2}\over{\cal M}_{\rm UV}^2}h_f^2
\quad \textrm{for}\ \ \textrm{Re}[\lambda]<0 \\ 
-\frac{\pi ^2 }{4\hbar}{\alpha_{\rm mode}^{3/2}b_2^{1/2}\over{\cal M}_{\rm UV}^2}h_f^2
\quad \textrm{for}\ \ \textrm{Re}[\lambda]>0\,. \end{cases}
\end{align}

We note that $S_{\rm on-shell}^{(2)}[N] $ depends on the lapse function $N$ only through the sign of $\textrm{Re}[\lambda]$. For instance, taking the tunneling saddle point $N_\textrm{T}$~\eqref{eq:tunneling-saddle}  
we have
\begin{align}\label{eq:tunneling-gcj-relation}
\lambda[N_\textrm{T}]=\frac{\sqrt{-\alpha_{\rm mode}\beta}N_\textrm{T}}{
(N^{2}_\textrm{T}H^{2}-q_f)^{2}}=-
\frac{\alpha_{\rm mode}^{3\over2}b_2^{1\over2}}
{4q_f{\cal M}_{\rm UV}^2}\left(1+i\sqrt{q_fH^2-1}\right),
\end{align}
which means $\textrm{Re}[\lambda]<0$.
Thus, we must set $C_3= 0$ and, as a result, we obtain the 
inverse Gaussian wave function for the tunneling proposal,
\begin{equation}
G[q_f, h_f] 
= G^{(0)}[q_f]\cdot \exp \left[{+\frac{\pi ^2 }{4\hbar}{\alpha_{\rm mode}^{3\over2}
b_2^{1\over2}\over{\cal M}_{\rm UV}^2}h_f^2}\right],
\end{equation}
meaning that the tensor perturbations lead to an unbounded distribution.
In contrast, the no-boundary saddle point $N_\textrm{H}$ takes $\textrm{Re}[\lambda]>0$
and leads to the Gaussian distribution for the tensor perturbations.
However, as previously discussed in GR,
the integration contours in the complex $N$ plane must pass through the tunneling saddle point $N_\textrm{T}$ even for the no-boundary proposal~\cite{Feldbrugge:2017mbc}.
Hence, even if the dispersion relation is modified 
as the generalized Corley-Jacobson dispersion 
relation~\eqref{Modified-dispersion-general} with $p=2$, 
the tensor perturbations exhibit inverse-Gaussian distribution for UV modes.
Although we have only obtained analytical solutions for $p=2$,
and make no analytical estimates for $p\neq2$, 
we numerically confirmed a similar behavior for $p\neq2$.
Although this analytical discussion was limited to the \ac{uv} region, Ref~\cite{Matsui:2022lfj} also numerically demonstrated that the perturbation problems occur in the general case as Eq.~\eqref{eq:eom-jc} where the dispersion relation takes the modified dispersion relation in the \ac{uv} region, and the standard form of \ac{gr} in the \ac{ir} region.

Next, we consider the open and flat universe. The sign of $\textrm{Re}[\lambda]$ determines 
the stability of the perturbations. For instance, taking the saddle point $N_\textrm{open}$~\eqref{eq:open-saddle},
we have 
\begin{align}\label{eq:open-gcj-relation}
\lambda[N_\textrm{open}]=\frac{\sqrt{-\alpha_{\rm mode}\beta}N_\textrm{open}}{
(N^{2}_\textrm{open}H^{2}-q_f)^{2}}=
\frac{\alpha_{\rm mode}^{3\over2}b_2^{1\over2}}
{4q_f{\cal M}_{\rm UV}^2}\frac{iH^2}{c_3+\sqrt{q_fH^2+1}},
\end{align}
which is imaginary. In the saddle point approximation, we can take either $C_3= 0$ or $C_4= 0$.
Thus, when we take $C_4= 0$, we can obtain the Gaussian wave function. 
On the other hand, taking the saddle point $N_\textrm{flat}^+$~\eqref{eq:flat-saddle}  we have
$\textrm{Re}[\lambda]>0$. Thus, we can take $C_4= 0$ and, as a result, we obtain the Gaussian wave function at the saddle-point approximation. 
Beyond the saddle-point approximation and considering the lapse integration over all complex planes, we obtain $\textrm{Re}[\lambda]<0$ again and must set $C_3= 0$ due to the regularity of the on-shell action.

Finally, we discuss the primordial inflation with the generalized Corley-Jacobson dispersion relation~\eqref{Modified-dispersion-general} with $p=2$. 
By using the background solution $q(t)=N^2H^2 t(t-1)+(q_f-q_i)t+q_i$, 
the perturbative solution reads  
\begin{align}\label{eq:gcj-infsolution}
\chi(t)=&-\frac{1}{2\sqrt{\alpha_{\rm mode}\beta} N} \Bigg[
((q_f + H^2 N^2 (t-1))t - t q_i + q_i) \times \nonumber \\
&\quad \exp \Big(
-i\sqrt{\alpha_{\rm mode}\beta} N \Big(
\frac{H^2 N^2 (2 t - 1)+q_f  - q_i}
{(q_f-q_i)^2 - 2 H^2 N^2 (q_f + q_i) + H^4 N^4}
\Big) \times \nonumber \\
&\quad \frac{-4 H^2 N^2 \tan^{-1}\Big(
\frac{H^2 N^2 (2 t - 1)+q_f  - q_i}
{\sqrt{(q_f-q_i)^2 - 2 H^2 N^2 (q_f + q_i) + H^4 N^4}}
\Big)}
{(2 H^2 N^2 (q_f + q_i) - (q_f-q_i)^2 - H^4 N^4)^{3/2}}
\Big) \times \nonumber \\
&\quad (2\sqrt{\alpha_{\rm mode}\beta} \tilde{C}_4 N \exp \Big(
2i\sqrt{\alpha_{\rm mode}\beta} N \Big(
\frac{H^2 N^2 (2 t - 1)+q_f  - q_i}
{(q_f-q_i)^2 - 2 H^2 N^2 (q_f + q_i) + H^4 N^4}
\Big) \times \nonumber \\
&\quad \frac{-4 H^2 N^2 \tan^{-1}\Big(
\frac{H^2 N^2 (2 t - 1)+q_f  - q_i}
{\sqrt{(q_f-q_i)^2 - 2 H^2 N^2 (q_f + q_i) + H^4 N^4}}
\Big)}{(2 H^2 N^2 (q_f + q_i) - (q_f-q_i)^2 - H^4 N^4)^{3/2}}\Big)\Big) - i \tilde{C}_3
\Bigg]\,, 
\end{align}
where $\tilde{C}_{3,4}$ are constants, 
When we assume $q_i\ll 1$, 
we approximate the above solution~\eqref{eq:gcj-infsolution} near $t = 0$ to be the above solution~\eqref{eq:gcj-solution} with $\tilde{C}_{3}\to C_3$ and $\tilde{C}_{4}\to C_4$.
Thus, the UV ($t=0$) contribution to the on-shell action with $\tilde{C}_{3}= 0$ vanishes for $\textrm{Re}[\lambda]<0$ whereas the contribution with $\tilde{C}_{4}= 0$ vanishes for $\textrm{Re}[\lambda]>0$. Other choices lead to a divergent on-shell action and we should not take other choices since it makes a large contribution to the on-shell action for $q_i\ll 1$. Thus, we reach the same conclusion with the quantum creation of the open universe. 
By imposing $\chi(1)=q_fh_f$ to normalize the overall factor for the solution~\eqref{eq:gcj-infsolution} with $\tilde{C}_3=0$ or $\tilde{C}_4=0$, we obtain
\begin{align}
\frac{i}{\hbar} S_\textrm{on-shell}^{(2)}[N]=\begin{cases} 
+\frac{\pi ^2 }{4\hbar}{\alpha_{\rm mode}^{3/2}b_2^{1/2}\over{\cal M}_{\rm UV}^2}h_f^2
\quad \textrm{for}\ \ \textrm{Re}[\lambda]<0 \\ 
-\frac{\pi ^2 }{4\hbar}{\alpha_{\rm mode}^{3/2}b_2^{1/2}\over{\cal M}_{\rm UV}^2}h_f^2
\quad \textrm{for}\ \ \textrm{Re}[\lambda]>0\,. \end{cases}
\end{align}
For the inflation case, we can take two real saddle points~\eqref{eq:inf-saddle}
and obtain, 
\begin{align}\label{eq:inf-gcj-relation}
\lambda[N_\textrm{inf}]=\frac{\sqrt{-\alpha_{\rm mode}\beta}N_\textrm{T}}{
(N^{2}_\textrm{T}H^{2}-q_f)^{2}}=i
\frac{\alpha_{\rm mode}^{3\over2}b_2^{1\over2}}
{H{\cal M}_{\rm UV}^2}\frac{(c_3q_{i}^{1/2}+q_{f}^{1/2})}{(q_{i}+2c_3q_{i}^{1/2}q_{f}^{1/2})^2},
\end{align}
which is imaginary. In the saddle point approximation, we can take either $\tilde{C}_3=0$ or $\tilde{C}_4=0$.
Thus, when we take $\tilde{C}_4=0$, we can obtain the Gaussian wave function
at two saddle points~\eqref{eq:inf-saddle}. 
In~\ac{gr} with saddle point approximation, the Gaussian perturbation is only achieved for the quantum creation of the flat universe, but it is also realized for the open universe and inflation model in the generalized Corley-Jacobson dispersion.

Although this behavior is in the \ac{uv} region, and in the \ac{ir} region the 
perturbation behavior is again the same with \ac{gr}, this result would be 
interesting when we consider the cosmological scenarios of \ac{hl} gravity. 
In the \ac{hl} gravity the \ac{uv} perturbations can produce scale-invariant primordial density perturbations and gravitational waves~\cite{Mukohyama:2009gg}.
Beyond the saddle-point approximation, it is necessary to set $\tilde{C}_3=0$ to 
ensure the regularity of the on-shell action, and the perturbation problem 
reappears. In conclusion, the issue of perturbations in Lorentzian quantum 
cosmology cannot be entirely resolved by considering trans-Planckian physics with 
the generalized Corley-Jacobson dispersion relation. However, the perturbations can be Gaussian in the quantum creation of the flat or open universe with the saddle-point approximation and the generalized Corley-Jacobson dispersion. 
This suggests that trans-Planckian physics or quantum gravity might offer a potential mechanism to stabilize perturbations in quantum cosmology.

%%%%%%%%%%%%%%%%%%%%%%%%%%%%%%%%%%%%
%%%%%%%%%%%%%%%%%%%%%%%%%%%%%%%%%%%%
\section{Conclusions and Discussions}
\label{sec:conclusions}
%%%%%%%%%%%%%%%%%%%%%%%%%%%%%%%%%%%%
%%%%%%%%%%%%%%%%%%%%%%%%%%%%%%%%%%%%

In this paper, we have investigated the perturbation problems in Lorentzian quantum cosmology, particularly focusing on various models of quantum cosmology where we have considered the closed, flat, and open universe. We have shown that this problem is inevitable as far as the regularity of the on-shell gravitational action is required, and in most quantum cosmological scenarios, the cosmological wave functions linearized perturbations exhibit an inverse Gaussian distribution and lead to cosmological inconsistencies.
At the saddle-point, the quantum creation of the flat universe has the Gaussian distribution, and this analysis corresponds to the standard inflation theory. However, beyond the saddle-point approximation and considering the $N$-integration over all complex planes, the perturbation problem appears again. Furthermore, we have considered the impact of trans-Planckian physics on quantum cosmology, using the generalized Corley-Jacobson dispersion relation as a
case study of modified dispersion relations. 
We have shown that even if we assume such a modified dispersion relation, the 
regularity of the on-shell action is required, and resolving perturbation problems 
in Lorentzian quantum cosmology remains a challenge. 
However, we have found that the perturbations can be Gaussian in the quantum 
creation of the open or flat universe and the primordial inflation  within the 
confines of the saddle-point approximation and the generalized Corley-Jacobson 
dispersion. This is because the behavior of the perturbation is changed by the modified dispersion relation, and therefore, the trans-Planckian physics or quantum gravity could solve the perturbation problem in quantum cosmology if the regularity of the on-shell action is improved or the perturbations in the \ac{uv} and \ac{ir} regions are modified in a closely related way.

%%%%%%%%%%%%%%%%%%%%%%%%%%%%%%%%%%%%%%%%%%%%%%%%%%%
\section*{Acknowledgment}
H.M. expresses gratitude to Shinji Mukohyama and Atsushi Naruko for collaborating on the earlier work~\cite{Matsui:2022lfj}. H.M. also thanks Masazumi Honda, Kazumasa Okabayashi, and Takahiro Terada for useful discussions regarding the Lefschetz thimble analysis. 
This work is supported by JSPS KAKENHI Grant No. JP22KJ1782 and No. JP23K13100.

%%%%%%%%%%%%%%%%%%%%%%%%%%%%%%%%%%%%
%%%%%%%%%%%%%%%%%%%%%%%%%%%%%%%%%%%%
\appendix
\section{Klein-Gordon norm and Bunch-Davies vacuum}
\label{appendix:Bunch-Davies}
%%%%%%%%%%%%%%%%%%%%%%%%%%%%%%%%%%%%
%%%%%%%%%%%%%%%%%%%%%%%%%%%%%%%%%%%%

To assist in understanding the perturbation problem in Lorentzian quantum cosmology~\cite{Feldbrugge:2017fcc,Feldbrugge:2017mbc}, and to explain why unstable perturbations occur in quantum cosmology, in this appendix we review the perturbation analysis of cosmic inflation. We will proceed with the path integral formulation for tensor perturbations in an asymptotically de Sitter background, taking the scale factor as $a(\eta) = -{1}/{H\eta}$ with the conformal time $-\infty < \eta < 0$.

Now, the second-order action for the tensor perturbation with wave-number $k$ around the spatially flat background spacetime is given as follows:
\begin{align}\label{Tensor-action}
\begin{split}
S_{\rm GR}^{(2)} &=2\pi^2 \int \mathrm{d}\eta 
\biggl[
\frac{a^2}{8}\left(h'\right)^2
-\frac{a^2}{8}k^2h^2\biggr]\\
&=2\pi^2 \int \mathrm{d}\eta 
\biggl[
\frac{1}{8}\left(\chi'\right)^2
-\frac{1}{8}\left( k^2-\frac{a''}{a}\right) \chi^2\biggr]
+\textrm{Boundary term}\,,
\end{split}
\end{align}
where we took $V_3=2\pi^2$, fixed the lapse function as $N=1$, introduced $\chi(\eta)=a(\eta) h(\eta)$, and used integration by parts. The equation of motion for $\chi(\eta)$ is as
\begin{align}\label{eq:eom1}
\chi''+\left(k^2-\frac{2}{\eta^2}\right) \chi=0\,.
\end{align}
We obtain a well-known solution,
\begin{align}\label{deSitter-conformal-solution}
\begin{split}
\chi(\eta)=\frac{A_k}{\sqrt{2k}}\left(1-\frac{i}{k\eta}\right)
e^{-ik\eta}+\frac{B_k}{\sqrt{2k}}\left(1+\frac{i}{k\eta}\right)e^{+ik\eta},
\end{split}
\end{align} 
where the Bunch-Davies vacuum corresponds to $B_k=0$.

We expand $\hat{\chi}(x)$ in terms of a set of complex solutions $\{\chi(x)\}$
of the Klein-Gordon equation~\eqref{deSitter-conformal-solution} as
\begin{equation}
\hat{\chi}(x) = \sum_k\left[\hat{a}_k\chi(x) + \hat{a}_k^{\dagger}\chi^*(x)
\right]
\end{equation}
where $\hat{a}_k$ and $\hat{a}_k^{\dagger}$ are the annihilation and creation operators 
associated with the set $\{\chi(x)\}$.
These operators satisfy the commutation relations, $[\hat{a}_k,\hat{a}_{k'}^{\dagger}]=\delta_{k{k'}}$ and 
$[\hat{a}_k,\hat{a}_{k'}]=[\hat{a}_k^{\dagger},\hat{a}_{k'}^{\dagger}]=0$.
The $\chi(x)$ should be complete and orthonormal 
concerning the Klein-Gordon inner product (see, e.g. Ref.~\cite{Crispino:2007eb}), 
\begin{equation}
\left(\chi_1, \chi_2 \right)_{\rm KG}\coloneqq -i \int
\left[\chi_1\partial_{\eta}\chi_2^*
-\chi_2^*\partial_{\eta}\chi_1\right] dx^3
\end{equation}
where $\chi_{1,2}$ are the Klein-Gordon solutions.
The Klein-Gordon inner product is constant for all time $\eta$
and can be easily verified by 
\begin{align}
\begin{split}
\partial_{\eta}\left(\chi_1,\chi_2\right)_{\rm KG}&=
\chi_1\partial_{\eta}\left(\partial_{\eta}\chi_2^*\right)
-\chi_2^*\partial_{\eta}\left(\partial_{\eta}\chi_1\right) =0.
\end{split}
\end{align}
The Klein-Gordon norm for the solution~\eqref{deSitter-conformal-solution}
is given by
\begin{equation}
\left(\chi,\chi\right)_{\rm KG}=|A_k|^2-|B_k|^2,
\end{equation}
where the positive or negative frequency modes respectively have positive or negative Klein-Gordon norms. Generally, a positive Klein-Gordon norm is required when defining a vacuum state in quantum field theory.

Hereafter, we will demonstrate that a positive norm $|A_k|^2-|B_k|^2>0$ leads to a Gaussian wave function, while a negative norm $|A_k|^2-|B_k|^2<0$ leads to an inverse Gaussian wave function in the path integral. For convenience, we replace the normalized parameters $A_k = (k^{3/2}h_f/iH) \alpha_k$ and $B_k = (k^{3/2}h_f/iH) \beta_k$ in~\eqref{deSitter-conformal-solution} to achieve $h(\eta_f) = h_f$ at the limit $\eta_f \to 0^{-}$. Then we can derive the following on-shell action,
\begin{align}\label{Tensor-on-shell-action1}
S_\textrm{on-shell}^{(2)}[h_f]
&=\frac{\pi^2}{4} \int_{-\infty}^{\eta_f} \mathrm{d}\eta 
\biggl[{a^2}\left(h'\right)^2-k^2{a^2}h^2\biggr]
={\pi^2\over 4 }\left[a^2h'h \right]_{-\infty}^{\eta_f} 
 \notag \\
&=\frac{\pi^2h_f^2k^2  \left(\alpha_ke^{-2 i k\eta_f}-\beta _k\right) 
\left[ (1 + i k \eta_f) \alpha_k + (-1 +i k\eta_f) \beta_k e^{2 i k\eta_f} \right]}{4H^2\eta_f} \\
&+\lim_{\eta_0\to -\infty}
-\frac{i\pi^2h_f^2k^3  \left(\alpha_ke^{-2 i k\eta_0} -\beta _k\right) 
\left(\alpha_k+\beta_ke^{2 i k\eta_0}\right)}{4H^2},
\notag
\end{align}
where the last term originates from the infinite path.
By evaluating the semiclassical exponent at the limit $\eta_f\to 0^{-}$ 
we obtain,
\begin{align}
G^{(2)}[h_f] &=
\begin{cases}
e^{-\frac{\pi^2}{4H^2}k^3h_f^2 \, 
-\frac{i\pi^2}{4H^2\eta_f}k^2h_f^2+\cdots}\quad (\alpha_k=1,\ \beta_k=0), \\ 
e^{+\frac{\pi^2}{4H^2}k^3h_f^2 \, 
+\frac{i\pi^2}{4H^2\eta_f}k^2h_f^2+\cdots}\quad (\alpha_k=0,\ \beta_k=1), 
\end{cases} 
\end{align}
where the former takes positive norm (Bunch-Davies vacuum),
and the wave function of the universe takes the form of Gaussian
and suppressed amplitude for the tensor perturbation modes. 
On the other hand, the latter takes a negative norm,
and increasing amplitude for the tensor perturbation modes.
As we saw in Section~\ref{sec:Perturbation-analysis}, since the lapse function $N$ is integrated into the complex plane, the last term in the Lorentz path integral diverges. Therefore it is not possible to take any perturbative solutions. As a consequence (a rather complicated discussion is on the way), we can not take positive norms and the Bunch-Davies vacuum.

%%%%%%%%%%%%%%%%%%%%%%%%%%%%%%%%%%%%%%%%%%%%%%%%%%%%%%%%%%%%%%%%%%%%%%%%%%%%
%%%%%%%%%%%%%%%%%%%%%%%%%%%%%%%%%%%%%%%%%%%%%%%%%%%%%%%%%%%%%%%%%%%%%%%%%%%%
\bibliographystyle{utphys}
\bibliography{Refs.bib}
%%%%%%%%%%%%%%%%%%%%%%%%%%%%%%%%%%%%%%%%%%%%%%%%%%%%%%%%%%%%%%%%%%%%%%%%%%%%
%%%%%%%%%%%%%%%%%%%%%%%%%%%%%%%%%%%%%%%%%%%%%%%%%%%%%%%%%%%%%%%%%%%%%%%%%%%%

\end{document}